\documentclass[%
 reprint,
 amsmath,amssymb,
 aps,
]{revtex4-2}

\usepackage{graphicx}% Include figure files
\usepackage{dcolumn}% Align table columns on decimal point
\usepackage{bm}% bold math
\usepackage{hyperref}% add hypertext capabilities
\usepackage{slashed}
\usepackage{subfloat}
\usepackage{pgfplots,subfigure}

\usepackage{geometry}
\definecolor{acsblue}{RGB}{17,76,139}

\begin{document}
\fontsize{8}{9}\selectfont
\preprint{APS/123-QED}

\title{Probing quantum criticality near the BTZ black hole horizon: Insights from coupled fermion-antifermion pairs}

\author{Abdullah Guvendi}
\email{abdullah.guvendi@erzurum.edu.tr (Corr. Author)}
\affiliation{Department of Basic Sciences, Erzurum Technical University, 25050, Erzurum, Türkiye}

\author{Omar Mustafa}
\email{omar.mustafa@emu.edu.tr}
\affiliation{Department of Physics, Eastern Mediterranean University, 99628, G. Magusa, north Cyprus, Mersin 10 - Türkiye}

\date{\today}

\begin{abstract}
\vspace{0.15cm}
\setlength{\parindent}{0pt}

{\footnotesize In this study, we analytically examine the behavior of a fermion-antifermion (\(f\overline{f}\)) pair near the horizon of a static BTZ black hole using a fully covariant two-body Dirac equation with a position-dependent mass, \(m \rightarrow m(r)\). This formulation leads to a set of four first-order equations that can be reduced to a second-order wave equation, enabling the analysis of gravitational effects on quantum interactions. Two mass modifications are considered: (i) \(m \rightarrow m - a/r\), representing an attractive Coulomb interaction, and (ii) \(m \rightarrow m - a/r + b r\), corresponding to a Cornell potential. For case (i), an exact analytical solution is obtained, while for case (ii), conditionally exact solutions involving biconfluent Heun functions are derived. For the lowest mode (\(n=0\)), the results indicate that real oscillations without energy loss occur when \(a > 0.25\) in scenario (i) and \(a > 0.75\) in scenario (ii), suggesting stable oscillatory behavior. When \(a < 0.25\) in scenario (i) or \(a < 0.75\) in scenario (ii), the state exhibits decay, indicating instability below these critical thresholds. At \(a = 0.25\) (scenario (i)) and \(a = 0.75\) (scenario (ii)), the system reaches a state where its evolution ceases over time. These findings provide insights into the stability conditions of fermion-antifermion pairs near the black hole horizon and may have relevance for determining critical coupling strengths in systems such as holographic superconductors. Furthermore, this work adopts an effective semi-classical quantum gravity approach, offering a practical framework for incorporating gravitational effects. However, a more complete description of the system would require a deeper understanding of quantum gravity beyond computational methods. The results presented here may contribute to further studies exploring the influence of strong gravitational fields on quantum systems.} 

\end{abstract}


\maketitle

%\tableofcontents

\section{Introduction}
\vspace{0.15cm}
\setlength{\parindent}{0pt}

Research on the interaction dynamics between two fermions dates back to early investigations that followed the introduction of the Dirac equation \cite{dirac}. Initially, the focus was on fermion-fermion systems, which prompted Breit to develop a two-body Dirac equation. This equation, involving Dirac matrices replacing velocities and incorporating a modified Darwin potential to handle inter-particle interactions, was effective in scenarios of weak coupling \cite{breit}. However, it struggled with long-range interactions and high velocities due to retardation effects. In response to these challenges, Bethe and Salpeter proposed an alternative based on quantum field theory \cite{bethe-salpeter}. Yet, their approach encountered difficulties concerning relative time, leading to exploration of alternatives like the instantaneous-interaction approximation. Barut took a different route by devising a Dirac-Coulomb type two-body equation in 3+1 dimensions, with the aim of achieving exact solvability. His method, derived from the action principle in quantum electrodynamics, incorporated retardation effects, precise spin algebra, and general electromagnetic potentials \cite{barut}. However, Barut's equation was found to be unsolvable for familiar systems such as one-electron atoms, resulting instead in a system of coupled second-order differential equations \cite{nuri}. Nevertheless, exact solutions were discovered for specific two-body systems in both three-dimensional flat spaces \cite{guvendi-epjc1} and curved spaces \cite{guvendi-plb1, guvendi-plb2, guvendi-podu, guvendi-btz, guvendi-npb1, semra, new-guvendi}. However, applying this covariant two-body Dirac equation to interacting fermions in a black hole background remains a formidable challenge, often necessitating numerical methods even for associated one-body fields to achieve results \cite{hassan}.

\vspace{0.15cm}
\setlength{\parindent}{0pt}

The fundamental significance of the near-horizon geometry of black holes in the context of the \(\text{AdS/CFT}\) correspondence has been rigorously established \cite{maldacena} (see also \cite{encodes}). Within this theoretical framework, analyzing the interaction dynamics of \( f\overline{f} \) pairs in the vicinity of a black hole horizon presents a compelling avenue for research. However, obtaining exact solutions to the fully covariant many-body Dirac equation governing coupled \( f\overline{f} \) pairs in the near-horizon background of a rotating black hole remains an exceedingly challenging problem. To date, no exact analytical solutions for such a system have been reported in the literature. On the other hand, the static Bañados-Teitelboim-Zanelli (BTZ) black hole \cite{BTZ,carlip}, a solution to Einstein's field equations in \( 2+1 \) dimensions with a negative cosmological constant (\(\Lambda = -1/\ell^2\)), shares several key features with black holes in \( 3+1 \) dimensions. However, it is confined both locally and asymptotically within \(2+1\)-dimensional anti-de Sitter (AdS) spacetime, setting it apart from asymptotically flat spacetimes and those with curvature singularities \cite{carlip}. The spacetime background describing the near-horizon region of the static BTZ black hole was comprehensively introduced in \cite{corichi}. This \(2+1\)-dimensional static spacetime provides a rigorous framework for analyzing the stability of composite systems composed of interacting particles. A thorough investigation of such coupled pairs under known interactions near this black hole's horizon offers valuable insights into the influence of particle-particle interactions on their evolution, as discussed in \cite{guvendi-btz}. Despite previous studies on the evolution of relativistic oscillators \cite{spin-0, spin-1, AA}, the exploration of a composite system involving interacting $f\overline{f}$ pairs within a static black hole background remains relatively limited \cite{guvendi-btz}. This paper aims to fill this gap by examining the effects of particle-particle interactions, including Coulomb and Cornell-type interaction potentials, on the evolution of $f\overline{f}$ pairs coupled near the horizon of the static BTZ black hole. This research is of considerable importance for identifying quantum critical points (QCPs) \cite{QCP1} (see also \cite{QCP2}) and their relevance to the field of holographic superconductivity \cite{QCP3}. In the context of fermion-fermion systems near the horizon of a static black hole, QCPs refer to the conditions under which a phase transition or distinct quantum behavior arises in the system. In the vicinity of the black hole's event horizon, the combination of strong gravitational and electromagnetic fields (if present) influences the dynamics of coupled pairs, potentially leading to the formation of a QCP where the system undergoes a phase transition. These points are characterized by singularities in physical observables, such as the energy of $f\overline{f}$ pairs (or their distribution function), marking critical thresholds for phenomena such as spontaneous pair creation, pair annihilation or significant changes in the structure of the quantum state (see also \cite{QCP1,QCP2,QCP3}). Analyzing these critical points requires a deep understanding of the interactions between gravity, gauge fields, and quantum field theory, particularly in the presence of the black hole's gravitational effects \cite{QCP1,QCP2,QCP3}. This necessitates employing techniques from quantum field theory in curved backgrounds \cite{QCP3}. On the other hand, the holographic principle \cite{sugg-3}, a concept in theoretical physics, posits that all information within a given space can be encoded by a theory defined on its boundary. At the heart of this principle lies the AdS/CFT correspondence, which suggests a connection between a specific type of string theory formulated in AdS space and a conformal field theory defined on the boundary of that space. This correspondence establishes a duality between a gravitational theory in higher dimensions and a gauge theory in one lower dimensions \cite{hol1,hol2,hol3}. Holographic superconductors, theoretical models utilizing the AdS/CFT correspondence, describe superconductivity within strongly coupled systems \cite{hol1,hol2,hol3,hol4,hol5,hol6,hol7,hol8,hol9,hol10,hol11}. In this framework, a superconductor is modeled using a black hole in higher-dimensional AdS space. Charged fields can condense near the black hole horizon, leading to the spontaneous breaking of a $U(1)$ symmetry in the dual field theory \cite{hol1,hol2,hol3}. This condensation resembles pair formation in conventional superconductors, resulting in the emergence of superconductivity in the boundary theory. This approach may offer insights into the behavior of high-temperature superconductors and other strongly correlated systems challenging to study using traditional methods. The holographic model reproduces several superconductivity features, such as the existence of a superconducting gap and the phase transition between normal and superconducting states \cite{QCP3}. The study of black holes often concentrates on the near-horizon geometry, where gravitational effects are exceptionally strong. This region provides insights into fundamental aspects of gravity and quantum mechanics, including the black hole information paradox and entropy. According to the holographic principle, the degrees of freedom responsible for black hole entropy are encoded on the horizon, contributing to the understanding of statistical mechanics concerning these micro states \cite{hol1,hol2,hol3}. Studies of the near-horizon also investigate quantum effects in strong gravitational fields, such as Hawking radiation \cite{H1,K1}, contributing to the broader goal of developing a consistent theory of quantum gravity \cite{sugg-2} (see also \cite{sugg-1}). The connection between QCPs, holographic superconductivity, and black hole near-horizon physics is rooted in the use of the AdS/CFT correspondence \cite{hol1,hol2,hol3,hol4,hol5,hol6,hol7,hol8,hol9,hol10,hol11}. By mapping the complex problem of superconductivity to a simpler gravitational problem in higher dimensions, insights into condensed matter systems can be gleaned through the mathematical tractability of black hole physics. Hence, the study of black hole near-horizons provides crucial insights into the fundamental nature of gravity and quantum mechanics within the same holographic framework \cite{hol1,hol2,hol3,hol4,hol5,hol6,hol7,hol8,hol9,hol10}, demonstrating the power of holography in bridging different physics domains.\\

\vspace{0.15cm}
\setlength{\parindent}{0pt}

In this paper, our attention is directed towards examining the evolution of a coupled $f\overline{f}$ pair in the near-horizon region of a static BTZ black hole \cite{corichi}. We utilize a fully covariant two-body Dirac equation within a position-dependent mass framework, as recently introduced in \cite{new-guvendi,AO}, to examine these systems and analyze the results in the context of QCPs and holographic superconductivity. The structure of this paper unfolds as follows: In section (\ref{sec2}), we lay out the generalized framework of the covariant two-body equation tailored for a coupled $f\overline{f}$ pair near the static BTZ black hole horizon. Within this framework, we derive a set of equations governing the dynamics of relative motion for the pair. Section (\ref{sec3.1}) introduces precise solutions for this specific system, with a particular focus on the \(m \rightarrow m - a/r\) adjustment (see also \cite{new-guvendi,QCP3,AO}). Moving on to section (\ref{sec3.2}), we explore an alternative approach by incorporating the Cornell-type potential, \(m \rightarrow m - a/r+br\). Here, we present conditionally exact solutions based on the biconfluent Heun functions. Finally, we summarize and discuss our findings in section (\ref{sec4}). It is important to note that throughout our analysis, we use units in which \( G = 1 = \hbar = c \).

\section{\label{sec2}{Generalized two-body Dirac equation}} \label{sec2}

In this section, we derive a comprehensive expression for the fully covariant two-body equation that governs the behavior of an \( f\overline{f} \) pair near the horizon of a static BTZ black hole, taking into account the spatial variation of the particle mass. To describe this scenario, we develop a set of coupled equations. First, we provide an overview of the near-horizon region of the BTZ black hole, as introduced in \cite{corichi}. The Euclidean solution of the static BTZ configuration, characterized by a positive mass \( M \), is described by the following line element \cite{corichi}:
\begin{equation}
ds^2 = \frac{r^2 - M\ell^2}{\ell^2} d\tilde{\tau}^2 + \frac{\ell^2}{r^2 - M\ell^2} dr^2 + r^2 d\theta^2. \label{eq1}
\end{equation}
Through the identifications \(\ell \rightarrow i\ell\) and \(\tilde{\tau} \rightarrow i\tilde{\tau}\), and subsequently transforming \(r \rightarrow T\) and \(\tilde{\tau} \rightarrow T\), we derive the following metric \cite{corichi}:
\begin{equation}
ds^2 = -\frac{\ell^2}{T^2 + M\ell^2} dT^2 + \frac{T^2 + M\ell^2}{\ell^2} dR^2 + T^2 d\theta^2. \label{eq2}
\end{equation}
Specifically, it serves as the dual to the static BTZ black hole solution \cite{corichi}. By considering \( M = -\alpha^2 \) and focusing on the parameter \(\alpha^2 \in (0,1]\), this solution, referred to as a 'particle' solution, exhibits conical singularity (similar to cosmic strings \cite{guvendi-plb1}), without the presence of a horizon. The deficit angle for this cone is \(\Omega = 2\pi(1 - \alpha) \). When \(\alpha = 1\), we have a singularity-free global \( AdS_3 \) spacetime. By utilizing established identifications within the spacetime (\ref{eq1}), one derives the following line element:
\begin{equation}
ds^2 = -\frac{\alpha^2\ell^2 - r^2}{\ell^2} d\tilde{\tau}^2 + \frac{\ell^2}{\alpha^2\ell^2 - r^2} dr^2 + r^2 d\theta^2. \label{eq3}
\end{equation}
This line element characterizes the geometry of a conical singularity located at \( r = 0 \) with a deficit angle denoted by \(\Omega\) (see \cite{corichi}). By introducing the variable \( \rho^2 = \ell^2 - r^2/\alpha^2 \) \cite{corichi}, one can derive a line element that describes the near-horizon region of the static BTZ black hole. Subsequently, by performing the transformations \(\tilde{\tau} \rightarrow t\) and \(\theta \rightarrow \phi\), the resulting metric can be rewritten, with the signature (\(+,-,-\)), as follows \cite{corichi,QCP3,AA}:
\begin{equation}
ds^2 \approx \frac{\alpha^2\rho^2}{\ell^2} dt^2 - d\rho^2 - \alpha^2\ell^2 d\phi^2. \label{eq4}
\end{equation}
Considering the line element (\ref{eq4}), the covariant metric tensor $g_{\mu\nu}$ is written as $g_{\mu\nu}=\textrm{diag}\left(\alpha^2\rho^2/\ell^2, -1, -\alpha^2\ell^2\right)$, with its contravariant form being $g^{\mu\nu}=\textrm{diag }\left(\frac{\ell^2}{\alpha^2\rho^2}, -1, -\frac{1}{\alpha^2\ell^2}\right)$, since $g^{\mu\nu}g_{\mu\nu}=\textbf{I}_{3}$. In the case of equal masses of two fermions in this near-horizon region, the covariant many-body Dirac equation can be expressed as follows \cite{guvendi-plb2,guvendi-btz,guvendi-npb1}:
\begin{eqnarray}
&\left\lbrace \mathcal{H}^{f} \otimes \gamma^{t^{\overline{f}}}+ \gamma^{t^{f}}\otimes \mathcal{H}^{\overline{f}} \right\rbrace \Psi\left(\textbf{x}_{1},\textbf{x}_{2} \right)=0,\nonumber\\
&\mathcal{H}^{f}=\left[ \slashed{\nabla}_{\mu}^{f} +im(x^{\mu^f}) \textbf{I}_{2} \right],\quad \mathcal{H}^{\overline{f}}=\left[ \slashed{\nabla}_{\mu}^{\overline{f}} +im(x^{\mu^{\overline{f}}}) \textbf{I}_{2}\right],\nonumber\\ 
&\slashed{\nabla}_{\mu}^{f}=\gamma^{\mu^{f}}\left(\partial_{\mu}^{f}-\Gamma_{\mu}^{f} \right),\quad \slashed{\nabla}_{\mu}^{\overline{f}}=\gamma^{\mu^{\overline{f}}}(\partial_{\mu}^{\overline{f}}-\Gamma_{\mu}^{\overline{f}}),  \label{tbe}
\end{eqnarray}
In this equation, Greek indices denote the coordinates within the curved background spacetime ($\mu=t,\rho,\phi$). The \(\gamma^{\mu}\) represent the generalized Dirac matrices, defined by the relation \(\gamma^{\mu} = e^{\mu}_{(k)} \gamma^{(k)}\), where \(e^{\mu}_{(k)}\) are the inverse tetrad fields and \(\gamma^{(k)}\) (with \(k = 0, 1, 2\)) are the spacetime-independent Dirac matrices. The spacetime-independent Dirac matrices are expressed through the Pauli spin matrices \((\sigma_{x}, \sigma_{y}, \sigma_{z})\) as follows: \(\gamma^{0} = \sigma_{z}\), \(\gamma^{1} = i\sigma_{x}\), \(\gamma^{2} = i\sigma_{y}\), for a (\(2+1\))-dimensional metric with negative signature, where the Minkowski tensor \(\eta_{(k)(l)}\) is given by \(\eta_{(k)(l)} = \text{diag}(+, -, -)\) \cite{guvendi-plb2,guvendi-btz,guvendi-npb1}. The inverse tetrad fields are determined using \(e^{\mu}_{(k)} = g^{\mu\tau} e_{\tau}^{(l)} \eta_{(k)(l)}\) \cite{spin-1}. Here, \(e_{\tau}^{(l)}\) represent the tetrad fields, obtained through \(g_{\mu\tau} = e_{\mu}^{(k)} e_{\tau}^{(l)} \eta_{(k)(l)}\). The spinorial affine connections, denoted by \(\Gamma_{\mu}\) in Eq. (\ref{tbe}), can be computed through \(\Gamma_{\lambda} = \frac{1}{8}\left[e^{(k)}_{\nu_{,\lambda}} e^{\tau}_{(k)} - \Gamma_{\nu\lambda}^{\tau}\right] \left[\gamma^{\mu}, \gamma^{\nu}\right]\), where \(\Gamma_{\nu\lambda}^{\tau}\) represent the Christoffel symbols, $\Gamma_{\nu \lambda}^{\tau} = \frac{1}{2} g^{\tau \epsilon} \left( \partial_{\nu} g_{\lambda \epsilon} + \partial_{\lambda} g_{\epsilon \nu} - \partial_{\epsilon} g_{\nu \lambda} \right)$ \cite{guvendi-plb2,guvendi-btz,guvendi-npb1}. Additionally, \(\textbf{I}_d\) denotes a d-dimensional identity matrix. Consequently, the following results can be obtained (see also \cite{spin-1}): 
\begin{equation}
\begin{split}
&e^{\mu^{f(\overline{f})}}_{(k)}=\textrm{diag}\left(\frac{\ell}{\alpha\rho^{f(\overline{f})}}, 1, \frac{1}{\alpha\ell} \right),\\
&\gamma^{t^{f(\overline{f})}}=\frac{\ell}{\alpha\rho^{f(\overline{f})}}\sigma_{z},\quad \gamma^{\rho^{f(\overline{f})}}=i\sigma_{x},\quad \gamma^{\phi^{f(\overline{f})}}=\frac{1}{\alpha\ell}i\sigma_{y},\\
&\Gamma_{t\rho}^{t^{f(\overline{f})}}=\frac{1}{\rho^{f(\overline{f})}}, \quad \Gamma_{t t}^{\rho^{f(\overline{f})}}=\frac{\alpha^2 \rho^{f(\overline{f})}}{\ell^2},\quad \Gamma_{t}^{^{f(\overline{f})}}=\frac{\alpha}{2\ell}\sigma_{y},\\
&\gamma^{\mu^{^{f(\overline{f})}}}\Gamma^{^{f(\overline{f})}}_{\mu}=-i\frac{1}{2\rho^{f(\overline{f})}}\sigma_{x}.\label{op}
\end{split}
\end{equation}
To provide further clarification, we can explicitly express the fully covariant two-body Dirac equation as \(\hat{M}\Psi = 0\), where \(\hat{M}\) is derived as follows:
\begin{eqnarray}
&\gamma^{t^{f}}\otimes\gamma^{t^{\overline{f}}}\left[\partial_{t}^{f}+\partial_{t}^{\overline{f}} \right]+im(x^{\mu^{f}},x^{\mu^{\overline{f}}})\left[\textbf{I}_{2}\otimes \gamma^{t^{\overline{f}}}+\gamma^{t^{f}}\otimes \textbf{I}_{2}\right]\nonumber\\
&+\gamma^{\rho^{f}}\partial_{\rho}^{f}\otimes \gamma^{t^{\overline{f}}}+ \gamma^{t^{f}}\otimes \gamma^{\rho^{\overline{f}}}\partial_{\rho}^{\overline{f}}\nonumber\\
&+\gamma^{\phi^{f}} \otimes\gamma^{t^{\overline{f}}}\partial_{\phi}^{f}+\gamma^{t^{f}}\otimes \gamma^{\phi^{\overline{f}}}\partial_{\phi}^{\overline{f}}\nonumber\\
&-\left[\gamma^{t^{f}}\Gamma_{t}^{f}\otimes \gamma^{t^{\overline{f}}}+\gamma^{t^{f}}\otimes \gamma^{t^{\overline{f}}}\Gamma_{t}^{\overline{f}} \right].   \label{exp1}
\end{eqnarray}
Considering the spacetime interval under scrutiny, we express the spacetime-dependent spinor \(\Psi(t,r,R)\) in a factorized manner as \(\Psi = e^{-i\omega t} \tilde{\Psi}(\vec{r},\vec{R})\). Here, \(\omega\) denotes the relativistic frequency, while \(\vec{r}\) and \(\vec{R}\) denote the spatial vectors for relative motion and center of mass motion, respectively. The corresponding coordinates are defined as outlined in \cite{guvendi-epjc1}
\begin{eqnarray*}
&r_{\mu} = x_{\mu}^{f} - x_{\mu}^{\overline{f}}, \quad R_{\mu} = \frac{x_{\mu}^{f} + x_{\mu}^{\overline{f}}}{2}, \quad x_{\mu}^{f} = \frac{1}{2}r_{\mu} + R_{\mu},\nonumber\\
&x_{\mu}^{\overline{f}} = -\frac{1}{2}r_{\mu} + R_{\mu}, \quad \partial_{x_{\mu}}^{f} = \partial_{r_{\mu}} + \frac{1}{2} \partial_{R_{\mu}},\nonumber\\
&\partial_{x_{\mu}}^{\overline{f}} = -\partial_{r_{\mu}} + \frac{1}{2} \partial_{R_{\mu}}, \label{eq4-}
\end{eqnarray*}
for two fermions of equal masses. It is important to emphasize that the combination \(\partial_{x_{\mu}}^{f} + \partial_{x_{\mu}}^{\overline{f}}\) equals \(\partial_{R_{\mu}}\). The evolution of the system, associated with the relativistic frequency \(\omega\), is delineated in terms of proper time. Our primary focus here lies in examining the relative motion of the $f\overline{f}$ pair, which is clarified under the assumption that the system's center of mass is stationary. With this premise, we can effectively decompose the spatial component of the spinor $\tilde{\Psi}(\rho,\phi)=e^{is\phi}(\psi_{1},\psi_{2},\psi_{3},\psi_{4})^{T}$, where $s$ denotes the total spin of the resultant composite system ($s=0,\pm 1$). As a result, we derive a set of coupled equations that govern the relative motion of the pair, considering that the total mass of the composite system depends on the relative radial coordinate ($m(\rho)$). By adding and subtracting these equations, we can rewrite the equation set in the following form, which includes two algebraic equations (see also \cite{guvendi-plb1,AO2})
\begin{equation}
\begin{split}
&\xi_{1}(\rho)-\frac{\tilde{s}}{\epsilon(\rho)}\,\xi_{3}(\rho)=0,\\
&\xi_{2}(\rho)-\frac{\hat{\lambda}}{\epsilon(\rho)}\,\xi_{3}(\rho)=0,\\
&\xi_{4}(\rho)-\frac{m(\rho)}{\epsilon(\rho)}\xi_{3}(\rho)=0,\\
&\epsilon(\rho)\xi_{3}(\rho)-m(\rho)\xi_{4}(\rho)+\hat{\lambda}\xi_{2}(\rho)-\tilde{s}\xi_{1}(\rho)=0,\label{eset}
\end{split}
\end{equation} 
where $\epsilon(\rho)=\frac{\omega \ell}{\alpha \rho}$, $\tilde{s}=\frac{s}{\alpha \ell}$, $\hat{\lambda}=\partial_{\rho}+\frac{1}{\rho}$, and 
\begin{equation*}
\begin{split}
&\xi_{1}(\rho)=\psi_{1}(\rho)+\psi_{4}(\rho),\quad \xi_{2}(\rho)=\psi_{1}(\rho)-\psi_{4}(\rho),\\
&\xi_{3}(\rho)=\psi_{2}(\rho)+\psi_{3}(\rho),\quad \xi_{4}(\rho)=\psi_{2}(\rho)-\psi_{3}(\rho).
\end{split}
\end{equation*} 
In Eq. (\ref{eset}), it can be observed that all other defined spinor components are expressible in terms of \(\xi_{3}(\rho)\). Consequently, the final equation in this set represents a second-order wave equation, which includes a general position-dependent mass term \(m(\rho)\).

\section{\label{sec3.1}{Exact solutions for \(m(\rho) \rightarrow m -\frac{a}{\rho}\)}} 

In this section, we present a precise solution for the system of equations outlined in Eq. (\ref{eset}) by adjusting the mass, which varies with position and is governed by the attractive Coulomb potential, \( -\frac{a}{\rho} \), where \( a \) represents the coupling constant. It is important to emphasize that the validity and effectiveness of the introduced model were demonstrated in Refs. \cite{new-guvendi,AO} for well-known \( f\overline{f} \) pairs, such as positronium, exciton, and quarkonium systems. Furthermore, this model can be generalized to include Cooper pairs without any loss of generality (see also \cite{NCom}). For this scenario, Eq. (\ref{eset}) leads to the following wave equation for \( \xi_3(\rho) \)
\begin{equation}
\xi_{3_{,\rho\rho}}+\frac{3}{\rho}\xi_{3_{,\rho}}+\left[\epsilon(\rho)^2-\left(m-\frac{a}{\rho}\right)^2-\tilde{s}^2+\frac{1}{\rho^2}\right]\xi_{3}=0,\label{WE1} 
\end{equation}
where, $_{,\rho}$ indicates derivative with respect to $\rho$. Introducing a new variable transformation, $z=2\sqrt{m^2+\tilde{s}^2}\,\rho$ (where $z\to 0 (\infty)$ as $\rho \to 0 (\infty)$), results in a wave equation that governs $\xi_{3}(z)$. Utilizing an ansatz function, $\xi_{3}(z)=z^{-3/2}\,\xi(z)$, simplifies this wave equation into a more familiar form
\begin{equation}
\left[\partial_{z}^2+\frac{\tilde{\mu}}{z}+\frac{\frac{1}{4}-\tilde{\nu}^2}{z^2}-\frac{1}{4} \right]\chi(z)=0, \label{DD}
\end{equation}
where 
\begin{equation*}
\tilde{\mu}=\frac{ma}{\sqrt{\tilde{s}^2+m^2}},\quad \tilde{\nu}=\sqrt{a^2-\tilde{\omega}^2},\quad \tilde{\omega}=\frac{\omega\ell}{\alpha}.
\end{equation*}
The regular solution to Eq. (\ref{DD}) around the regular singular point \( z = 0 \) can be expressed using the confluent hypergeometric function of the first kind, denoted as $\xi(z)=e^{-\frac{z}{2}}\,z^{\frac{1}{2}+\tilde{\nu}}\  _1F_{1}(\frac{1}{2}+\tilde{\nu}-\tilde{\mu}, 1+2\tilde{\nu}; z)$. It is crucial to highlight that we are focusing on a system confined within the near-horizon region, where the corresponding wave function(s) need to diminish as the spatial coordinate extends infinitely, denoted as $\xi(z)\propto e^{-z}$. However, it is acknowledged that the asymptotic behavior of the function $_1F_{1}(\frac{1}{2}+\tilde{\nu}-\tilde{\mu}, 1+2\tilde{\nu}; z)$ 
\begin{eqnarray*}
\approx  \frac{\Gamma(1+2\tilde{\nu})}{\Gamma(\frac{1}{2}+\tilde{\nu}-\tilde{\mu})}\frac{e^{z}}{z^{(\frac{1}{2}+\tilde{\nu}+\tilde{\mu})}}\left[1+\mathcal{O}(|z|^{-1}) \right], 
\end{eqnarray*}
divergent when $z\to\infty$ \cite{guvendi-btz}. In our pursuit of a well-behaved solution to the wave equation, we aim to define it using the condition: \(\frac{1}{2} + \tilde{\nu} - \tilde{\mu} = -n\) where \(n\) represents the radial quantum number \((n = 0, 1, 2, \dots)\). This condition forms the foundation for deriving the relativistic frequency expression relevant to the system under investigation
\begin{equation}
\omega_{ns}=-i\frac{\alpha}{\ell}\sqrt{a^2\left[\frac{m^2}{(m^2-\tilde{s}^2)}-1\right]+\tilde{n}^2-\frac{2am\tilde{n}}{\sqrt{m^2+\tilde{s}^2}}}\label{spec}
\end{equation}
in which $\tilde{n}=n+\frac{1}{2}$. When $s=0$, this expression becomes
\begin{equation}
\omega_{n}=-i\frac{\alpha}{\ell}\sqrt{\tilde{n}^2-2a\tilde{n}}\label{n-states}.
\end{equation}
When \(a\) approaches zero, it becomes evident that the lowest mode (\(n=0\)) becomes \(\omega \to -i\alpha/\ell\), which gives the formal Hawking temperature of the background, a finding consistent with earlier research outlined in \cite{corichi,QCP3}. This lowest mode can manifest as a real oscillation, particularly noticeable when \(a > 0.25\), with \(\omega_{0} = \frac{\alpha}{\ell} \sqrt{a - 1/4}\). This suggests that, in principle, such a composite system decays over time if \(a < 0.25\). We can expect the system to eventually converge to the ground state. This state can be a stable quantum state if \(a > 0.25\). Moreover, there is a notable exception when \(a = 0.25\), which results in a unique quantum state, that is a QCP, where the system cannot exist over time, since \(\omega_{0} = 0\) in this unique state (see also \cite{QCP3}). In Eq. (\ref{spec}), by taking $m=0$ (see also \cite{AO}), the frequency expression becomes
\begin{equation}
\omega_{n}=-i\frac{\alpha}{\ell}\sqrt{\tilde{n}^2-a^2}\label{massless-case}.
\end{equation}
For a coupled massless pair in the ground state, we determine new quantum critical points (QCPs) at \( a = 0.5 \) for \( f\overline{f} \) systems and \( a = -0.5 \) for the corresponding \( ff \) systems, such as the bielectron system \cite{NCom}.

\section{\label{sec3.2}{Conditionally exact solutions in case where \(m(\rho) \rightarrow m -\frac{a}{\rho}+b\rho\)}} 

In this section, we introduce a solution that is conditionally precise for the equations outlined in Eq. (\ref{eset}) by adjusting the position-dependent mass in accordance with the Cornell-type interaction potential, \(-\frac{a}{\rho}+b\rho\) \cite{AO}. In this setting, Eq. (\ref{eset}) leads to the subsequent wave equation:
\begin{equation*}
\xi_{3_{,\rho\rho}} + \frac{3}{\rho}\xi_{3_{,\rho}} + \left[\epsilon(\rho)^2 - \left(m - \frac{a}{\rho}+b\rho\right)^2 - \tilde{s}^2 +\frac{1}{\rho^2}\right]\xi_{3} = 0. \label{WE2}
\end{equation*}
This equation can be written as follows:
\begin{equation}
\xi_{3_{,\rho\rho}} + \frac{3}{\rho}\xi_{3_{,\rho}}+\left[\frac{\eta_{1}}{\rho^2} -\eta_{2}+\frac{\eta_{3}}{\rho} -b\rho\left(2m+b\rho\right)\right]\xi_{3} = 0, \label{WE2.1}
\end{equation}
and its solution around the origin (regular singular point) \cite{AO} can be represented as
\begin{equation*}
\xi_{3}(\rho)=\mathcal{C}\exp(-\zeta)\, \rho^{\sigma}\,\mathcal{H_{B}}\left(\tilde{\alpha}, \tilde{\beta}, \tilde{\gamma}, \tilde{\delta};\, y\right),
\end{equation*}
where $\mathcal{C}$ denotes an arbitrary constant, $y=\sqrt{b}\rho$
\begin{equation}
\begin{split}
&\zeta=m\rho+\frac{b\rho^2}{2},\quad \sigma=-1+\sqrt{1-\eta_{1}},\\
&\eta_{1}=\tilde{\omega}^2-a^2+1,\quad \eta_{2}= m^2+\tilde{s}^2-2ab,\quad \eta_{3}=2ma, \\
&\tilde{\alpha}=2\sqrt{a^2-\tilde{\omega}^2},\quad \tilde{\beta}=\frac{2m}{\sqrt{b}},\quad \tilde{\gamma}=\frac{2ab-\tilde{s}^2}{b},\\
&\tilde{\delta}=-\frac{4am}{\sqrt{b}}.   \label{par}
\end{split}
\end{equation}
It is important to acknowledge that we deliberately neglected the second part of the radial function $\xi_{3}(\rho)$. This omission is due to its association with $\sigma=-1-\sqrt{1-\eta_{1}}$, which results in an infinite solution at $y=0$. Additionally, the biconfluent Heun function/series $\mathcal{H}_{B}(\tilde{\alpha},\tilde{\beta},\tilde{\gamma},\tilde{\delta}; y)$ needs to be truncated to a polynomial of order $n_r+1\geq 1$. This requirement is outlined in the detailed new approach presented in the Appendix of Mustafa \cite{omar-1}, and applied by Mustafa and collaborators \cite{omar-2,omar-3}. Specifically, this approach mandates truncating the biconfluent Heun functions/series to a polynomial of order $n_r+1\geq 1$ under two conditions:
\begin{equation}
\tilde{\gamma}=2(n_r+1)+\tilde{\alpha},\label{FC}
\end{equation} 
and
\begin{equation}
\tilde{\delta}=-\tilde{\beta}(2n_r+\tilde{\alpha}+3).\label{SC} 
\end{equation}
One should observe that the first condition is the standard biconfluent Heun polynomials whereas the second condition was very recently introduced ( for more details see \cite{omar-1,omar-2,omar-3} and related references cited therein)  to facilitate conditional exact solvability of the system at hand.  These conditions suggest that $b\rightarrow\tilde{s}^2$, and accordingly we can determine the frequency expression as the following 
\begin{equation}
\omega_{n}= -i\frac{\alpha}{\ell}\sqrt{n^{2}_{*}-2an_{*}},\quad n_{*}=n+\frac{3}{2},
\end{equation}
for such a coupled pair in the near-horizon of the static BTZ black hole. When the system reaches its ground state, there are three possible scenarios. Firstly, if \( a < 0.75 \), the system is unstable and decays over time (for more details see \cite{QCP3}). Secondly, if \( a > 0.75 \), the system undergoes real oscillations with a frequency \(\omega_{0} = \frac{\alpha}{\ell}\sqrt{3a - 9/4}\). In this case, the system is stable and does not decay over time, as there is no indication of energy loss (note that \(\Psi \propto \textrm{e}^{-i\omega t}\)). Lastly, a notable exception occurs when \( a = 0.75 \), leading to a QCP where the system cannot exist over time since \(\omega_{0} = 0\) in this state.    

\section{Summary and discussions}\label{sec4}

In this manuscript, we investigate the evolution of a $f\overline{f}$ pair near the horizon of a static BTZ black hole. To analyze this system, we use a fully-covariant two-body Dirac equation with a position-dependent mass setting, \(m \rightarrow m(\rho)\). Initially, we derive a set of four first-order equations that collectively form a second-order wave equation. We then explore two distinct scenarios for the mass modification: (i) \(m \rightarrow m - a/\rho\), representing an attractive Coulomb interaction, and (ii) \(m \rightarrow m - a/\rho + b \rho\), corresponding to a Cornell potential function. For the first scenario, we demonstrate that an exact solution is attainable. For the second scenario, we find solutions using conditionally exact solutions of the biconfluent Heun functions. We discuss various cases within the first and second scenarios. We observe that for the lowest mode (\(n=0\)), a real oscillation can occur without any indication of energy loss, particularly when \(a > 0.25\) and \(s=0\), or when \(a > 0.5\) and \(m=0\) in the first scenario, or when \(a > 0.75\) in the second scenario. In contrast, the mode decays over time if \(a < 0.25\) and \(s=0\), or if \(a < 0.5\) and \(m=0\) in the first scenario, or if \(a < 0.75\) in the second scenario. Notably, when \(a = 0.25\) and \(s = 0\), or when \(a = 0.5\) and \(m = 0\) in the first scenario, or when \(a = 0.75\) in the second scenario, the system reaches unique states where it cannot persist or evolve over time. 

\vspace{0.15cm}
\setlength{\parindent}{0pt}

In the first scenario, the corresponding modes decay over time with decay time $\tau_{ns}=\frac{1}{|\Im \omega_{ns}|}$ \cite{QCP3,AO2}
\begin{equation}
\tau_{ns}=\frac{\ell}{\alpha\sqrt{a^2\left[\frac{m^2}{(m^2-\tilde{s}^2)}-1\right]+\tilde{n}^2-\frac{2am\tilde{n}}{\sqrt{m^2+\tilde{s}^2}}}}\label{dt1}
\end{equation}
and relatively long lived modes seem possible when $\alpha<<1$. This decay time expression becomes as the following 
\begin{equation}
\tau_{n}=\frac{\ell}{\alpha\sqrt{\tilde{n}^2-2a\tilde{n}}},\label{dt2}
\end{equation}
for singlet quantum state ($s=0$) of such a coupled pair. This suggests that the decay time of the system in its ground state (\(n=0\)) tends to infinity as \(a \to 0.25\), provided that \(a < 0.25\). Moreover, decay time of these states can be altered by the parameters $\alpha$ and $\ell$. The same discussion also applies when \(a \to 0.5\) and \(m=0\) in the first scenario. 

\vspace{0.15cm}
\setlength{\parindent}{0pt}

In the second scenario, the corresponding modes decay over time with a lifetime: 
\begin{equation}
\tau_{n}= \frac{\ell }{\alpha \sqrt{n^{2}_{*}-2an_{*}}},\quad n_{*}=n+\frac{3}{2}.\label{dt3}
\end{equation}
In this case, the decay time of a system in its ground state tends to infinity as \(a \to 0.75\), provided that \(a < 0.75\), and the decay rates are influenced by the parameters \(\alpha\) and \(\ell\). Our results indicate that the evolution of coupled $f\overline{f}$ pairs depends not only on particle-particle interactions but also on the spacetime parameters. 

\vspace{0.15cm}
\setlength{\parindent}{0pt}

The QCP is a fundamental concept in condensed matter physics, representing a zero-temperature (or zero-frequency) point where a continuous phase transition occurs, driven by quantum fluctuations rather than thermal fluctuations, which dominate at higher temperatures. In our initial scenario(s), we examined coupled pairs interacting through the Coulomb potential and determined critical coupling values corresponding to situations where information about energy and time evolution is lost (see also \cite{QCP3}). Below these critical values, the corresponding systems cannot sustain a stationary state, with its decay rate depending on spacetime parameters and the coupling strength (\(\propto a\)) between the particles. For coupling strengths exceeding \(a > 0.25\), the spinless pair supports a stable mode, with the ground state frequency given by \(\omega = \frac{\alpha}{\ell} \sqrt{a - 1/4}\). When a Cornell-type coupling is introduced, the QCP shifts to \(a = 0.75\) for such a coupled pair in the ground state. These results imply the presence of a phase transition at the QCPs. Recently, QCPs were also studied in the context of charged scalar fields with position-dependent mass near the horizon of a static BTZ black hole. By comparing our findings with those in \cite{QCP3}, it can be concluded that the singlet quantum state of coupled $f\overline{f}$ pairs is effectively described by solutions to the corresponding Klein-Gordon equation (see also \cite{guvendi-epjc1}). This also applies to massless cases. Accordingly, we can also conclude that the linear term in the Cornell potential has an analogous effect to an external magnetic field, shifting the ground state to the first excited state in an effective manner \cite{QCP3}.

\vspace{0.15cm}
\setlength{\parindent}{0pt}

In the systems analyzed in Sections \ref{sec3.1} and \ref{sec3.2}, QCPs are identified, where the field transitions to a non-propagating state at zero frequency or energy. At these QCPs, the field frequencies drop to zero, indicating a change from stable oscillatory behavior to a non-propagating regime. Below these critical points, the system exhibits instability, characterized by negative purely imaginary frequencies that signify rapid decay and high dissipation. Above the critical points, the modes stabilize and propagate, marking a transition to a superconducting-like phase with the disappearance of dissipation and the dominance of stable excitations (see \cite{QCP3} for further details). The appearance of negative purely imaginary frequencies is closely connected to the modified temperature, which approaches zero as the system reaches the QCPs. At these critical points, the black hole stops emitting thermal radiation, signaling a transition to a zero-temperature phase associated with a quantum phase transition. In this phase, the system is characterized by purely quantum states, where quantum effects dominate. The Hawking temperature, which governs the thermal radiation emitted by black holes due to quantum fluctuations at the event horizon, also influences their evaporation. In the context of holographic superconductivity, this temperature plays a crucial role through the AdS/CFT correspondence, where the temperature of the black hole’s horizon corresponds to that of the boundary theory. As the system cools toward the zero temperature, it transitions from a normal state to a superconducting phase, resembling the behavior of conventional superconductors with zero electrical resistance.

\vspace{0.15cm}
\setlength{\parindent}{0pt}

Our findings offer a refined perspective on the stability of fermion-antifermion pairs near a black hole horizon, providing insights into the critical coupling strengths associated with the formation of stable pairs in superconductors, particularly within the framework of holographic superconductivity. Additionally, this study presents a semi-classical quantum gravity approach as a useful framework for investigating such systems. However, a more complete description would require a deeper understanding of quantum gravity, extending beyond computational and algorithmic methods \cite{sugg-1,sugg-2,sugg-3}. While this work provides an effective approximation, it also establishes a foundation for future studies incorporating quantum gravity principles.

\section*{Credit authorship contribution statement}

\textbf{Abdullah Guvendi}: Conceptualization, Methodology, Formal Analysis, Writing – Original Draft, Investigation, Visualization, Writing – Review and Editing.\\
\textbf{Omar Mustafa}: Conceptualization, Methodology, Formal Analysis,  Writing – Original Draft, 
Investigation, Writing – Review and Editing. 

\section*{Data availability}

This manuscript has no associated data.

\section*{Conflicts of interest statement}

No conflict of interest declared by the authors.

\section*{Funding}

No fund has received for this study.

%\bibliography{apssamp}% Produces the bibliography via BibTeX.

\end{document}